\documentclass[aps,pra,a4paper,preprint,showpacs]{revtex4}
\usepackage{epsfig}
\usepackage{graphicx}
\usepackage{amsmath}
\usepackage{bm}
\newcommand{\be}[1]{\begin{equation}\label{#1}}
\newcommand{\ee}{\end{equation}}
\newcommand{\bea}{\begin{eqnarray}}
\newcommand{\eea}{\end{eqnarray}}

\newcommand{\ud}{\mathrm{d}}
\newcommand{\re}{\mathrm{Re}}

\begin{document}
\title{Semiclassical initial value calculations of the collinear helium atom}
\author{C Harabati}
\email{harabac@mail.biu.ac.il}
\author{K G Kay}
\email{kay@mail.biu.ac.il}
\affiliation{Department of Chemistry, Bar-Ilan University, Ramat-Gan,
Israel 52900} 
\date{\today}

\begin{abstract}
Semiclassical calculations using the Herman-Kluk initial value
treatment are performed to determine energy eigenvalues of bound and
resonance states of the collinear helium atom. Both the $eZe$
configuration (where the classical motion is fully chaotic) and the
$Zee$ configuration (where the classical dynamics is nearly
integrable) are treated. The classical motion is regularized to remove
singularities that occur when the electrons collide with the
nucleus. Very good agreement is obtained with quantum energies for
bound and resonance states calculated by the complex rotation method.
\end{abstract}
\maketitle
\section{Introduction}

The inability of the old quantum theory to determine the energy
spectrum of helium was one of the factors motivating the development
of modern quantum mechanics. With the advent of the modern theory,
attempts to treat helium within a classical framework continued with
the application of semiclassical methods, obtained from an asymptotic
analysis of the Schr\"odinger equation.\cite{Tanner00}. These studies
eventually identified two key reasons for the failure of old theory's
attempts to describe the energy levels of this atom: the absence of
the Maslov index in the older work,\cite{Leopold80} and the largely
chaotic nature of the classical electronic dynamics in this
system.\cite{Richter90,Blumel92} Due to this last factor, truly
satisfactory semiclassical treatments had to await development of
techniques for the quantization of chaotic motion.  The first modern
semiclassical calculations to properly account for the chaotic
dynamics in helium appeared in the early
1990's.\cite{Ezra91,Wintgen92} These and subsequent works have enabled
the quantization of a large number of bound and resonance states of
the atom.\cite{Tanner00,Ezra91,Wintgen92,Rost97,Wintgen94}

Although other restricted-configuration coplanar studies have been
carried out,\cite{Grujic95} most recent semiclassical treatments are
based on a collinear model for the classical motion of the electrons,
with corrections added for stable motion in the remaining degrees of
freedom.\cite{Tanner00} Collinear helium can exist in two
configurations:
\begin{itemize}
\item The $eZe$ configuration, characterized by the
two electrons on different sides of the nucleus. This configuration is
relevant for quantization of bound and resonant states of the
three-dimensional atom with maximal positive values of the Stark
quantum number $K$ including, apparently, the ground state of the
atom. The classical motion for this configuration is believed to be
completely chaotic, i.e., the periodic orbits are all linearly
unstable and proliferate exponentially with orbit
length. Semiclassical quantization in this case was performed by a
periodic orbit cycle expansion
technique.\cite{Ezra91,Wintgen92,Rost97}

\item The $Zee$ configuration, characterized by the
two electrons on the same side of the nucleus. This configuration is
implicated in the quantization of certain resonance states with
maximal negative values of $K$. In strong contrast to the $eZe$ case,
the classical behavior for the $Zee$ configuration is nearly
integrable, with invariant tori centered about a stable periodic orbit
describing the so-called "frozen planet" motion.\cite{Richter92b}
Quantization for this case was achieved by applying the
Einstein-Brillouin-Keller (EBK) treatment.\cite{Wintgen94,Richter90b,Richter91}
\end{itemize}

The justification for treating helium with reduced dimensionality is
that many highly excited quantum states are dominated by the classical
collinear motion. This conclusion rests on three
arguments:\cite{Tanner00} (a) A scaling analysis shows that the
classical dynamics of helium with total orbital angular momentum ${\bf
L}$ at energy $E$ is equivalent to the dynamics of the system with
angular momentum $|E|^{1/2}\mathbf{L}$ at energy $-1$. Thus, as the
energy approaches the double ionization limit $E=0$, the classical
motion relevant for states with low and moderate angular momentum
quantum numbers becomes that for ${\bf L}=0$, which restricts the
three particles to a plane.  (b) It is known that the collinear $eZe$
and $Zee$ configurations form invariant subspaces in the
full-dimensional phase space. (c) It can be shown that the collinear
motion is stable with respect to bending deformations perpendicular to
the linear configuration. On the basis of these arguments many states
of the three-dimensional atom are expected to be structured about the
backbone of collinear classical motion. This reduction in
dimensionality was crucial for the ability to quantize states
associated with the $eZe$ configuration since it enabled a thorough
study and coding of the system's periodic orbits.

Although the existing semiclassical treatments quite successfully
reproduce energy levels for many states of the three-dimensional atom,
full-dimensional calculations of helium are highly desirable.  One
problem with the collinear treatments is that they are applicable only
to states with near-extreme values of the Stark quantum number; they
are not truly appropriate for more general states.  A second
limitation of these approaches is they are not strictly justified for
low-lying excited states of the atom where the arguments favoring
collinearity, or even coplanarity, break down. This comment applies
even to $S$ states which should, arguably, be treated semiclassically
with nonzero $\mathbf{L}$ due to the Langer
correction.\cite{Manning94} One consequence is that the accuracy of
the semiclassical approximation for the ground state of helium has not
yet been definitively established.  Unfortunately, the semiclassical
methods that have been used to quantize the chaotic classical motion
for collinear $eZe$ helium cannot be easily applied to
higher-dimensional systems where the dynamics generally have a mixed
chaotic-regular nature and where a systematic search and
characterization of all periodic orbits becomes unfeasible. Clearly,
this limitation also blocks extension of these semiclassical
techniques to atomic and molecular systems having more than two
electrons. Thus, it is important to develop other semiclassical
approaches that are capable of treating higher-dimensional
many-electron systems.

Semiclassical initial value representation (IVR)
methods\cite{Herman84,Kluk86,Miller06,Thoss04,Kay05} are natural
candidates for this role since they are capable of semiclassically
propagating wave functions for systems with many degrees of
freedom. Here we test the applicability of one such method for the
quantization of many-body Coulombic systems by applying it to the
collinear helium atom. In contrast to much of the previous
semiclassical work for this system, no corrections for the motion
perpendicular to the linear configuration are incorporated in the
present treatment, so that the results apply to the collinear system
only, not to the three-dimensional atom.

Although IVR treatments have been previously applied to
single-electron systems\cite{Kay05,VandeSand99,Yoshida04,Takatsuka06},
the present application to a system with two electrons is nontrivial.
The most serious difficulty concerns the strongly chaotic nature of
the classical motion for the $eZe$ configuration. It is well
established that such behavior creates numerical convergence
difficulties in the propagation of systems by IVR methods beyond very
short times.\cite{Kay94c,Walton96,Spanner05} Thus, calculations for
times long enough to extract quantization information are expected to
be challenging.  A related problem is associated with the classical
instability of the helium atom with respect to ionization. In contrast
to the quantum system, the classical atom can autoionize at all
energies, even those below the first ionization limit. Thus, in
general, one electron of the atom escapes to infinity after a few
revolutions, leaving a bound He$^+$ ion.  This is an extreme example
of the zero-point energy problem\cite{Miller89,Bowman89} that plagues
classical trajectory simulations of quantum evolution.  More
specifically, the potential difficulty in the present context arises
from the reliance of IVR methods on general classical trajectories
that are almost never true periodic orbits. Quantization is obtained
from those trajectories that shadow the periodic orbits for long time
intervals. The short lifetime of many trajectories, resulting from the
classical autoionization, may require a drastic increase in the total
number of trajectories that must be calculated. A final problem
anticipated for IVR treatments of the collinear atom in both
configurations (and one that is common to all numerical semiclassical
calculations for these systems) arises because the electrons undergo
repeated collisions with the nucleus, where the Coulombic interaction
becomes infinite.  At such collisions, the classical equations of
motion become singular, and they cannot be integrated by ordinary
methods.  Numerical solution of the classical equations of motion for
collinear helium thus requires regularization techniques.

The remainder of this paper is organized as follows. Section II
presents the expressions that form the basis of the present
treatment. Section III describes the semiclassical calculations, Sec.\
IV presents the numerical results, and Sec.\ V discusses and
summarizes the main points of our work. An Appendix presents details
of supporting quantum calculations for the energy levels of the
collinear atom.

\section{Theory}

In atomic units, the Hamiltonian for the collinear helium atom is
given by
\begin{equation}\label{ham}
H=\frac{p_1^2}{2}+\frac{p_2^2}{2}-\frac{2}{q_1}-\frac{2}{q_2}+\frac{1}{|q_2\pm
q_1|},
\end{equation}
where ${\bf q}^T=(q_1, q_2)$ are the coordinates and ${\bf p}^T=(p_1,
p_2)$ are the momenta of the two electrons. The $+$ and $-$ signs in
the last term apply to the $eZe$ and $Zee$ configurations,
respectively.

Here we wish to determine the energies of bound and resonance states
for these two configurations. In a purely quantum calculation, such
energies can be extracted from peaks in the spectrum
\begin{equation}\label{ft}
F(E) = \left |\int_0^{\infty}\, \exp(iEt)\, c(t)\, dt \right |^2
\end{equation}
where $c(t)$ is the autocorrelation function
\begin{equation} \label{autoc}
c(t) =  \int d\mathbf{r}'\int d\mathbf{r}\, \Psi_0^{*}(\mathbf{r}')
K(\mathbf{r}',\mathbf{r},t)  \Psi_0(\mathbf{r}).
\end{equation}
Here  $\Psi_0$ denotes an initial state which projects onto energy
eigenstates of interest and
\begin{equation}
K(\mathbf{r}',\mathbf{r},t) = \langle
\mathbf{r}'|\exp(-i\hat{H}t)| \mathbf{r}\rangle
\end{equation}
is the propagator, describing the time evolution operator in the
representation of positions ${\bf r}^T=(r_1,r_2)$. In the present
semiclassical calculations, this propagator is approximated
semiclassically.

The specific semiclassical treatment used here is that of Herman and
Kluk (HK)\cite{Herman84,Kluk86} who expressed the propagator in the
IVR form
\begin{equation}\label{hkp}
K^{HK}({\bf r}',{\bf r}, t)=\int\!\!\!\int\frac{\ud^2q\,
\ud^2p}{(2\pi)^2}\,\langle {\bf r}'|{\bf q}_t,{\bf p}_t\rangle\,
R_{{\bf q}{\bf p}t} \,
\exp (iS_{{\bf q}{\bf p}t})\,\langle {\bf q},{\bf p}|{\bf r}\rangle,
\end{equation}
where the integrals are over momenta $\mathbf{p}$ and coordinates
$\mathbf{q}$ that serve as initial conditions for classical
trajectories. The trajectories carry these phase space variables at
time 0 to values $(\mathbf{p}_t,\mathbf{q}_t)$ at time $t$. The
quantities
\begin{equation}\label{gwp}
\langle {\bf r} | {\bf q},{\bf p}\rangle=
\left(\frac{\textrm{det}\bm{\gamma}}{\pi^2}\right)^{1/4}\exp
\left[-\frac{1}{2}({\bf r}-{\bf q})^T\bm{\gamma} ({\bf r}-{\bf q})+
i{\bf p}^T({\bf r}-{\bf q})\right] 
\end{equation}
are Gaussian coherent state functions. $\bm{\gamma}$ is a $2\times 2$
diagonal matrix with real nonzero elements $\gamma_1,\gamma_2$ which
determine the Gaussian widths along $r_1$ and $r_2$. The function
$S_{{\bf q}{\bf p}t}=\int_0^t \ud t'
(\mathbf{p}_{t'}\dot{\mathbf{q}}_{t'} -H )$ is the action along the
trajectory and the prefactor $R_{{\bf q}{\bf p}t}$ has the form
\begin{equation}\label{pfa}
R_{{\bf q}{\bf p}t}=\left \{{\textrm{det}\left [ \frac{1}{2}\left(\frac{\partial
{\bf q}_t}{\partial {\bf q}}+
\bm{\gamma}^{-1}\frac{\partial {\bf p}_t}{\partial {\bf
p}}\bm{\gamma}-i\frac{\partial {\bf q}_t}{\partial {\bf
p}}\bm{\gamma} 
+i \bm{\gamma}^{-1}\frac{\partial {\bf p}_t}{\partial {\bf
q}}\right) \right ]} \right \}^{1/2}, 
\end{equation}
where each term under the radical is a $2\times 2$ block of the
4-dimensional monodromy matrix defined by
\begin{equation}\label{sma}
\left( \begin{array}{c}
\delta {\bf q}_t \\
\delta {\bf p}_t
\end{array} \right)
=\left(\begin{array}{cc}
\partial {\bf q}_t/\partial {\bf q} & \partial {\bf q}_t/\partial {\bf p} \\
\partial {\bf p}_t/\partial {\bf q} & \partial {\bf p}_t/\partial {\bf p}
\end{array} \right)
\left( \begin{array}{c}
\delta {\bf q} \\
\delta {\bf p}
\end{array} \right).
\end{equation}

A potential difficulty with applying the HK propagator in the present
case is that (\ref{hkp}) is technically valid only for systems
with wave functions that obey boundary conditions at $r_i = \pm
\infty$. In contrast, the wave functions of collinear helium are
required to obey boundary conditions at $r_i=0$ (where they must
vanish\cite{Blumel92}).  Due to the Gaussian form of the functions
$\langle {\bf r}' | {\bf q}_t,{\bf p}_t\rangle$, the wave functions
calculated using the HK treatment will not satisfy this property.  To
remedy this problem, the IVR expression for the propagator should be
modified to replace the Gaussian coherent state functions with certain
functions that vanish at $r_i=0$ and obey some additional
requirements.\cite{Kay05,Kay01}.  Fortunately, however, these
modifications are not needed for the present calculations. The states
$\Psi_0(\mathbf{r})$, which we use to prepare the system at time 0 and
to probe the positions of the particles at times $t$, are localized in
regions far from $r_i=0$. Thus, the initial states effectively obey
the proper boundary conditions and the function $c(t)$ is insensitive
to the behavior of the propagated wave functions
$\Psi_t(\mathbf{r}')=\int d\mathbf{r}
K^{HK}(\mathbf{r'},\mathbf{r},t)\Psi_0(\mathbf{r})$ near $r_i'=0$. The
results of an HK calculation should then be the same as for a modified
IVR propagator in which the coherent state functions are replaced by
new functions that are strongly non-Gaussian at $r_i=0$ (as needed to
obey the boundary conditions) but become almost Gaussian far from the
boundaries. This was confirmed by calculations in which the $\langle
{\bf r} | {\bf q},{\bf p}\rangle$ and $\langle {\bf r}' | {\bf
q}_t,{\bf p}_t\rangle$ were replaced by combinations of Gaussian
coherent state functions which vanish at $r_i=0$. The results were
essentially identical to those obtained with the ordinary HK
expression.

As mentioned above, it is necessary to regularize the classical
equations of motion because they become singular when electrons
collide with the nucleus ($q_{1t}=0$ or $q_{2t}=0$) and cannot be
integrated by applying ordinary numerical techniques.  As a prelude to
this regularization, it is convenient to transform the coordinates and
momenta to scaled variables defined by
\begin{equation}\label{scale}
\tilde{q}_i = |E|q_i, \qquad \tilde{p}_i=
|E|^{-1/2}p_i \qquad (i=1,2)
\end{equation}
where $E$ is the energy of the atom.  In terms of these variables, the
Hamiltonian can be expressed as $H=|E|\tilde{H}$ where
\begin{equation}
\tilde{H}
=\frac{\tilde{p}_1^2}{2}+\frac{\tilde{p}_2^2}{2}-\frac{2}{\tilde{q}_1}-\frac{2}{\tilde{q}_2}+\frac{1}{|\tilde{q}_2\pm
\tilde{q}_1|} 
\end{equation}
has an energy-independent value. For the present treatment of bound
and resonance states (for which $E<0$) we set $\tilde{H}=-1$.  The
physical variables $(\mathbf{q}_t,\mathbf{p}_t)$ at time $t$ can be
calculated by solving Hamilton's equations of motion with Hamiltonian
$\tilde{H}$ to obtain the scaled variables at the time $\tilde{t}$,
defined by
\begin{equation} \label{tscale}
\tilde{t} = |E|^{3/2}t,
\end{equation}
and then applying (\ref{scale}).
 
To complete the regularization procedure, we apply a transformation
similar to that proposed by Kustaanheimo and
Stiefel\cite{Kustaanheimo65,Aarseth74}. In our case, this involves
defining a new Hamiltonian $\mathcal{H} =
\tilde{q}_1\tilde{q}_2(\tilde{H}+1)$ which is numerically equal to
zero and is free from singularities. Introducing the regularized
variables $(Q_i,P_i)$, defined via the equations
\begin{equation}\label{catr}
\tilde{q}_i = Q_i^2 \qquad
\tilde{p}_i = \frac{P_i}{2Q_i}\qquad (i=1,2),
\end{equation}
$\mathcal{H}$ takes the form
\begin{equation}\label{reham}
\mathcal{H}=\frac{(Q_2^2P_1^2+Q_1^2P_2^2)}{8}-2(Q_1^2+Q_2^2)+Q_1^2Q_2^2
+\frac{Q_1^2Q_2^2}{|Q_2^2\pm Q_1^2|}.
\end{equation}
With this Hamiltonian, the classical equations of motion for
$dQ_{i\tau}/d\tau$ and $dP_{i\tau}/d\tau$ are solved with initial
conditions $(Q_i,P_i)$ at time 0 to determine values of
$(Q_{i\tau},P_{i\tau})$ as a function of a new time variable
$\tau$. This variable is related to the scaled time $\tilde{t}$ by the
differential equation
\begin{equation}\label{titr}
\frac{d\tilde{t}}{d\tau} = Q_{1\tau}^2Q_{2\tau}^2,
\end{equation}
which must be integrated along with Hamilton's equations.  The
physical variables $p_{it},q_{it}$ at time $t$ are then obtained by
reversing the regularization and scaling transformations.

The physical action integral $S_t$ needed to form the HK integrand can
be obtained from the regularized variables by applying 
\begin{equation}
S_t = |E|^{-1/2}{\mathcal S}_{\tau},
\end{equation}
where the scaled action ${\mathcal S}_{\tau}$ is determined by integrating
the equation
\begin{equation}\label{act}
\frac{d{\mathcal
S}_{\tau}}{d\tau}=\frac{(Q_{2\tau}^2P_{1\tau}^2+Q_{1\tau}^2P_{2\tau}^2)}{8} 
+2(Q_{1\tau}^2+Q_{2\tau}^2)
-\frac{Q_{1\tau}^2Q_{2\tau}^2}{|Q_{2\tau}^2\pm Q_{1\tau}^2|}.
\end{equation}

To evaluate the monodromy elements needed for the semiclassical
calculation, we begin with the relations
\begin{eqnarray}
|E|\, q_{it} & = & Q_{i\tau}^2, \nonumber\\
2 \,Q_{i\tau}\, p_{it} & = & \sqrt{|E|}\,\, P_{i\tau}, 
\end{eqnarray}
which are obtained by combining  (\ref{scale}) and (\ref{catr}).
Treating $(Q_{i\tau}, P_{i\tau})$ as functions of $t$ and the initial
conditions $\alpha = \{q_1,q_2,p_1,p_2\}$, and taking partial
derivatives with respect to any one of these variables, $\alpha_j$, we
obtain
\begin{eqnarray} \label{transf1}
|E|\,\frac{\partial q_{it}}{\partial \alpha_j} & = &
2\,Q_{i\tau}\,\left(\frac{\partial Q_{i\tau}}{\partial
\alpha_j}\right)_t-q_{it}\frac{\partial |E|}{\partial
\alpha_j}, \nonumber\\ 
Q_{i\tau}\,\frac{\partial p_{it}}{\partial \alpha_j} & = &
-p_{it}\,\left(\frac{\partial Q_{i\tau}}{\partial
\alpha_j}\right)_t+\frac{\sqrt{|E|}}{2}\,\left(\frac{\partial
P_{i\tau}}{\partial \alpha_j}\right)_t+\frac{P_{i\tau}}{4\sqrt{
|E|}}\,\frac{\partial |E|}{\partial \alpha_j}. 
\end{eqnarray}
To determine the partial derivatives of $Q_{i\tau}$ and $P_{i\tau}$
with respect to $\alpha_j$ appearing above, we treat $Q_{i\tau}$ and
$P_{i\tau}$ as functions of the initial values $Q_i,P_i$ and of
$\tau$. Differentiating with respect to $\alpha_j$ and applying the
equations of motion then gives
\begin{eqnarray}
\left(\frac{\partial Q_{i\tau}}{\partial \alpha_j}\right)_t & = &
\frac{\partial Q_{i\tau}}{\partial Q_{k}}\,\frac{\partial
Q_{k}}{\partial \alpha_j}+\frac{\partial Q_{i\tau}}{\partial
P_{k}}\,\frac{\partial P_{k}}{\partial \alpha_j}+\frac{\partial
\mathcal{H}}{\partial P_{i\tau}}\,\left(\frac{\partial \tau}{\partial
\alpha_j}\right)_t,\nonumber\\ 
\left(\frac{\partial P_{i\tau}}{\partial \alpha_j}\right)_t & = &
\frac{\partial P_{i\tau}}{\partial Q_{k}}\,\frac{\partial
Q_{k}}{\partial \alpha_j}+\frac{\partial P_{i\tau}}{\partial
P_{k}}\,\frac{\partial P_{k}}{\partial \alpha_j}-\frac{\partial
\mathcal{H}}{\partial Q_{i\tau}}\,\left(\frac{\partial \tau}{\partial
\alpha_j}\right)_t. 
\end{eqnarray}
Finally, to determine $(\partial \tau/\partial \alpha)_t$ in the above
equations, we differentiate $\tilde{t}$ (considered as a function of
initial conditions and $t$) with respect to $\alpha_j$ and apply 
(\ref{tscale}) and (\ref{titr}). This yields
\begin{equation}\label{tidri}
\frac{3}{2}\,\sqrt{|E|}\,\frac{\partial |E|}{\partial
\alpha_j}\,t=\frac{\partial \tilde{t}}{\partial Q_{k}}\,\frac{\partial
Q_{k}}{\partial \alpha_j}+\frac{\partial \tilde{t}}{\partial
P_{k}}\,\frac{\partial P_{k}}{\partial
\alpha_j}+Q_{1\tau}^2\,Q_{2\tau}^2\,\left(\frac{\partial \tau}{\partial
\alpha_j}\right)_t, 
\end{equation}
from which $(\partial \tau/\partial \alpha)_t$ can be obtained.  These
equations thus express the physical monodromy matrix in terms of the
regularized monodromy matrix and the partial derivatives $\partial
\tilde{t}/\partial Q_k$ and $\partial \tilde{t}/\partial P_k$. The
various quantities needed to calculate the physical monodromy matrix
can now be formed by solving a set of first-order ordinary
differential equations in the variable $\tau$, obtained by
differentiating Hamilton's equations (for $dQ_{i\tau}/d\tau$ and
$dP_{i\tau}/d\tau$) and (\ref{titr}) (for $d
\tilde{t}/d\tau$) with respect to $Q_{k}$ and $P_{k}$.

We mention that an alternative procedure for calculating the physical
monodromy matrix can be formulated in terms of a transformation to
so-called local coordinates\cite{Richter93}. However, the present
approach was adopted here since numerical tests showed that it allows
larger time steps for the integration of the differential equations.

Note that the differential equations that are solved here remain well
behaved for all values of $\tau$. The singularities in the physical
variables (monodromy elements and momenta), occurring at collisions
with the nucleus, arise only from (\ref{transf1})-(\ref{tidri}),
describing transformations from the regularized variables.
Nevertheless, such singularities cause the prefactor $R_{{\bf q}{\bf
p}t}$ to behave as $q_{it}^{-1}$ for $q_{it}\to 0$, raising the
suspicion that the HK integral diverges for the present
system. Fortunately, these singularities in $R_{{\bf q}{\bf p}t}$ are
cancelled by factors such as $\exp[-(p_{it}-p_{i0})^2/4\gamma_i]$
appearing in the expressions for $c(t)$ [see (\ref{ovlp}) below],
which cause the HK integrand to vanish when $q_{it}\to 0$ since
$p_{it}\propto q_{it}^{-1/2}$ in such cases.

While the regularization presented above provides a unique
continuation of the classical equations of motion when one of the
coordinates $q_{1t}$ or $q_{2t}$ becomes zero (a binary collision), no
unique continuation exists when both $q_{1t}$ and $q_{2t}$ become zero
simultaneously (a triple collision). Fortunately, the initial
collisions leading to an exact triple collision form a set of zero
measure in the system's phase space and do not contribute to
calculations of the HK propagator.

\section{semiclassical calculations}

The symmetry of the Hamiltonian in (\ref{ham}) implies that two
distinct kinds of energy eigenstates exist: states that are symmetric
with respect to electron exchange (which can be identified with
singlet spin states) and states that are antisymmetric with respect to
electron exchange (which can be identified with triplet spin states).
As in the case of exact quantum evolution, one expects that energy
spectra for the two kinds of states can be obtained separately by
semiclassically propagating initial states of definite exchange
symmetry. Thus, in our calculations, we choose the functions $\Psi_0$
in (\ref{autoc}) to have the (anti)symmetrized form
\begin{equation} 
\Psi_0(\mathbf{r}) = {\mathcal N}\left [\langle r_1 | q_{10}, p_{10}\rangle  \langle r_2
| q_{20}, p_{20}\rangle \pm \langle r_1 | q_{20}, p_{20}\rangle  \langle r_2
| q_{10}, p_{10}\rangle \right ],
\end{equation}
where $ \langle r_i | q_{j0}, p_{j0}\rangle$ are one-dimensional
Gaussian coherent state functions involving parameters
$(q_{j0},p_{j0})$ and ${\mathcal N}$ is a normalization constant. We
choose the values of $\gamma_j$ for these functions to be equal to
those used to construct the functions $\langle {\bf r} | {\bf q},{\bf
p}\rangle=\langle r_1 | q_{1}, p_{1}\rangle\langle r_2 | q_{2},
p_{2}\rangle $ in the propagator.  However, rather than propagate a
combination of coherent state functions, it is more convenient to
propagate a single, unsymmetrized, ``initial'' state and project out
the desired symmetry component with a properly symmetrized
``final'' state.\cite{Kay94c} Applied to the autocorrelation function
[(\ref{autoc})], this treatment yields
\begin{eqnarray}\label{autoc1}
c^{HK}(t) =2 {\mathcal N}^2\int\!\!\!\int \frac{\ud^2q\,
\ud^2p}{(2\pi)^2} &\left [\langle q_{10},p_{10}|q_{1t},p_{1t}\rangle
\langle q_{20},p_{20}|q_{2t},p_{2t}\rangle\pm \langle
q_{20},p_{20}|q_{1t},p_{1t}\rangle \langle
q_{10},p_{10}|q_{2t},p_{2t}\rangle \right]\nonumber \\ & R_{{\bf
q}{\bf p}t}
\exp(iS_{{\bf q}{\bf p}t})
\langle q_1, p_1 | q_{10},p_{10}\rangle\langle q_2,p_2|q_{20},p_{20}\rangle.
\end{eqnarray}
 
The product of overlap factors on the extreme right in 
(\ref{autoc1}) can be expressed in the analytical form
\begin{eqnarray} \label{ovlp}
\langle {\bf q},{\bf p} |{\bf q}_0,{\bf p}_0\rangle
=&\exp\left[-\frac{1}{4}({\bf q}-{\bf q}_0)^T\gamma({\bf q}-{\bf q}_0)
-\frac{1}{4}({\bf p}-{\bf p}_0)^T\gamma^{-1}({\bf p}-{\bf
p}_0)\right]\nonumber \\ 
&\times\exp\left[\frac{i}{2}({\bf p}+{\bf p}_0)^T({\bf q}-{\bf
q}_0)\right].
\end{eqnarray}
In our calculations, we used the Gaussian distribution, formed from
the modulus of this expression, to sample points in phase space for a
Monte Carlo evaluation of the integrals over $(\mathbf{q},\mathbf{p})$
in (\ref{autoc1}).

To calculate the integrand in  (\ref{autoc1}), the transformations
(\ref{scale}) and (\ref{catr}) were applied to determine values for
the regularized variables $Q_{i}, P_{i}$ for each physical point
$(\mathbf{q},\mathbf{p})$ sampled. Using these as initial conditions,
Hamilton's equations for $Q_{i\tau}$ and $P_{i\tau}$ were solved
numerically, along with (\ref{titr}) and (\ref{act}), as
well as the $\tau$-dependent differential equations for elements
[e.g. $\partial Q_{i\tau}/\partial Q_{j}$] of the regularized
monodromy matrix and the quantities $\partial\tilde{t}/\partial
Q_{i}$, $\partial\tilde{t}/\partial P_{i}$.  This set of 26 equations
was integrated using a variable stepsize Runge-Kutta
method\cite{Press92}. The values of $\tau$, corresponding to an
evenly-spaced sequence of desired times $t$, were determined by a
root-solving procedure and, at such times, the transformations were
reversed to obtain the physical coordinates, momenta, action, and
monodromy elements, needed for construction the integrand.

The final step in the calculation, namely, the evaluation of the
Fourier transform in (\ref{ft}), was performed by the harmonic
inversion procedure,\cite{Wall95,Mandelshtam97a,Mandelshtam97b} which
effectively fits the computed autocorrelation function to an
expression of the form
\begin{equation}\label{harminv}
c^{HK}(t)=\sum_k d_k \exp(-i E_kt),
\end{equation} 
extracting values for the complex parameters $E_k$ and $d_k$.  This
method was preferred over ordinary Fourier transform techniques since
it was capable of producing spectra of higher resolution spectra from
our autocorrelation data, which were calculated for relatively short
time durations.

Numerical difficulties, caused by the highly chaotic nature of the
classical dynamics for the $eZe$ configuration, made it necessary to
apply some additional approximations.  As is well known, the chaotic
motion causes the modulus of the HK prefactor $R_{{\bf q}{\bf p}t}$ to
increase approximately exponentially with
time.\cite{Kay94c,Walton96,Spanner05} This systematic increase is
distinct from singularities in the prefactor that occur when an
electron collides with the nucleus.  In principle, chaos should also
cause the phases in the HK integrand to vary rapidly as a function of
initial condition and the resulting cancellations should compensate
for the growth of the prefactor, reducing the autocorrelation function
to physically correct values. However, as time progresses, the
prefactor becomes so large that the number of Monte Carlo-sampled
trajectories needed for this cancellation exceeds practical limits.
As a result, the calculation cannot be made to converge beyond a short
time. The practice of discarding trajectories with large
prefactors\cite{Kay94c,Spanner05} does not, by itself, solve this
problem for the present system since, to achieve convergence, it is
necessary to eliminate almost all trajectories that do not autoionize.

To reduce the severity of this difficulty we observe that it is the
phase of the prefactor, rather than its modulus, that is primarily
responsible for creating the interference in the phase space
integration which results in quantization. Thus, as other works
demonstrate,\cite{Kaledin03a,Kaledin03b,Takatsuka07} it is possible to
improve numerical convergence of the autocorrelation expression
(without adversely affecting the computed energies) by replacing the
modulus of the prefactor with a less rapidly increasing function of
time, while retaining the accurate value of the phase. The choice for
this replacement is not very critical, but a constant value for
$|R_{{\bf q}{\bf p}t}|$ is found not to work well here since the
rapidly varying integrand phases then cause the resulting
autocorrelation signal to decay too quickly. In the present work, we
have found it useful to replace the computed value of $|R_{{\bf q}{\bf
p}t}|$ with $|R_{{\bf q}{\bf p}t}|^{1/2}$.

Despite this step, we still find it necessary to discard a very small
minority of trajectories (e.g., about one trajectory out of $10^5$)
having extremely large prefactors (e.g., $|R_{{\bf q}{\bf
p}t}|>10^{20}$) which would otherwise destroy the numerical
convergence of the calculations.  We emphasize that the above steps
were used only for the calculations of the $eZe$ configuration. For
calculations of the nonchaotic $Zee$ configuration, the correct HK
expressions for the prefactors were used and no trajectories were
discarded due to large $|R_{{\bf q}{\bf p}t}|$,

One final simplification adopted here was to filter out trajectories
with energies sufficiently far from the average energy of the initial
state $\Psi_0$ and the target energy eigenvalues.  Thus, trajectories
with positive energies (above the double ionization limit) and
sufficiently negative energies were eliminated from the phase space
integrations. Removal of trajectories with low energies was especially
important for the $eZe$ configuration since they become highly
unstable at relatively short times, leading to very large prefactors,
and making it difficult to achieve convergence for the required time
durations.

\section{results}

As in the case of the three dimensional helium atom, the $eZe$ and
$Zee$ collinear systems have bound, resonance, and continuum states.
We denote bound and resonance states with the symbol $N_n$ where
$N=1,2, \ldots$ and $n=N,N+1, \ldots$ are, respectively, approximate
quantum numbers for the inner electron and outer electrons. For each
$N$, the states $N_n$ form a Rydberg series converging, as $n \to
\infty$, to an ionization threshold with energy $E=-2/N^2$~a.u.  The
series associated with $N=1$ ($E<-2.0$ a.u.) are bound states while
those associated with $N>1$ are resonances.

\begin{figure}
\begin{center}
\includegraphics[scale=0.65]{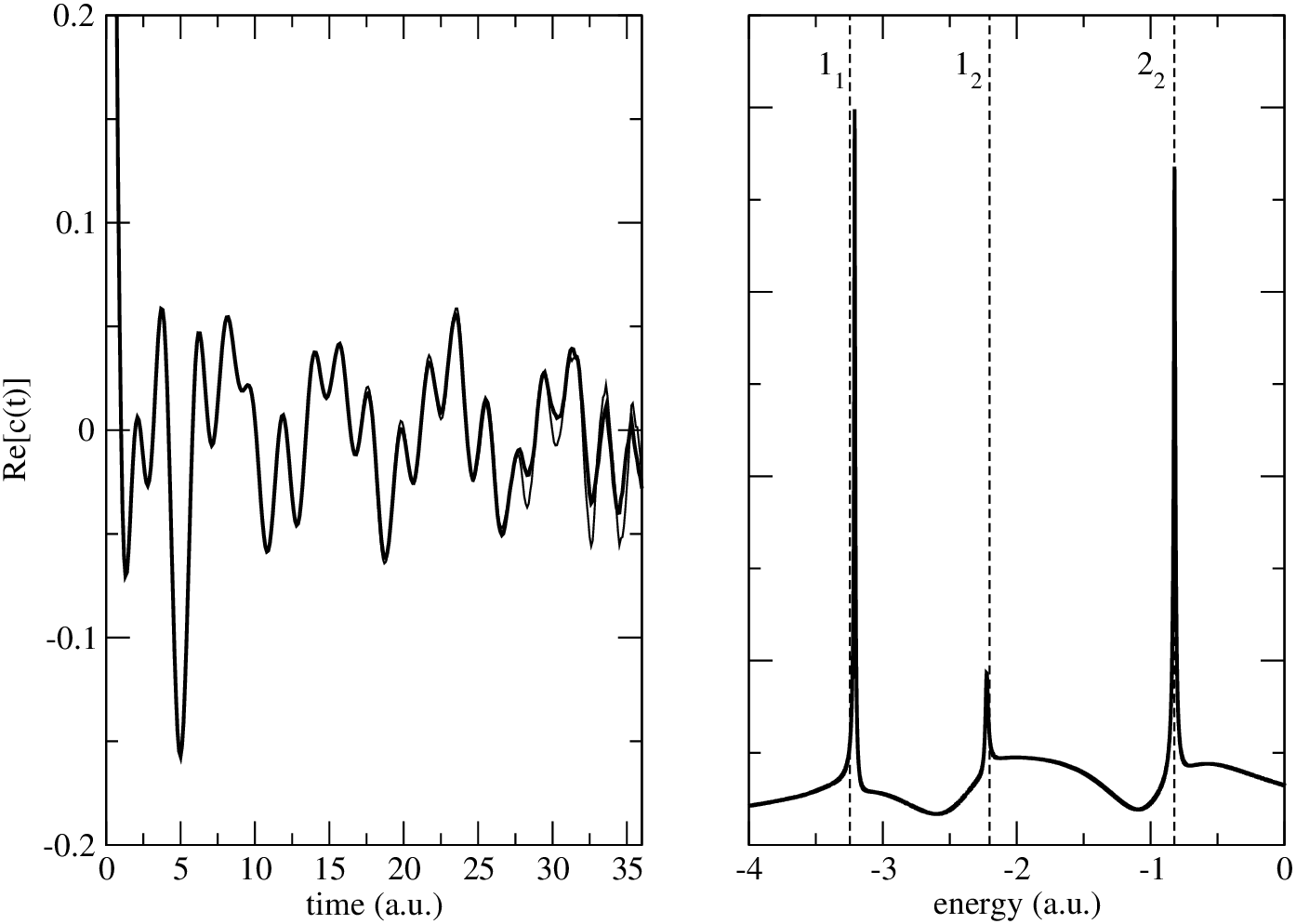}
\caption{\label{fig:grouspe}
Results obtained for the $eZe$ configuration with the initial state
having parameters ${\bf q}_0^T=(2.0,2.0)$, ${\bf p}_0^T=(0,0)$ and
$\gamma_1=\gamma_2=1.0$ a.u. The left panel shows the real part of
autocorrelation functions calculated using about $2.3\times 10^7$
trajectories (heavy curve) and $1.1\times 10^7$ trajectories (light
curve). The right panel shows the energy spectrum obtained by the
harmonic inversion method. The vertical dashed lines indicate quantum levels.}
\end{center}
\end{figure}

Figure \ref{fig:grouspe} shows the autocorrelation function and
spectrum obtained for the $eZe$ configuration using an initial state
with ${\bf q}_0^T=(2.0,2.0)$, ${\bf p}_0^T=(0,0)$ and
$\gamma_1=\gamma_2=1.0$. This same state was previously used by
Simonovi\'c\cite{Simonovic04} in a quantum mechanical calculation of
the autocorrelation function for this system. The spectrum obtained
from that calculation had three visible peaks, corresponding to
energies of the ground state $1_1$, the first excited state $1_2$,
and the lowest resonance state $2_2$.  The present semiclassical
calculations with this initial state were performed using $2.3\times
10^{7}$ trajectories having energies between -3.5 and 0 a.u.  The
convergence of the resulting autocorrelation function can be judged by
comparing the heavy curve in the left panel of figure \ref{fig:grouspe}
with the light curve, which was obtained using half the number of
trajectories. The harmonic inversion spectrum shown in the right panel
of figure \ref{fig:grouspe} has peaks for the same three levels
observed in the split propagator quantum
calculation.\cite{Simonovic04} Also shown in the figure are vertical
lines denoting quantum mechanical energies, computed by the complex
rotation method (see Appendix A for details). The
semiclassical-quantum agreement is seen to be good, even for the
ground state

\begin{figure}
\begin{center}
\includegraphics[scale=0.65]{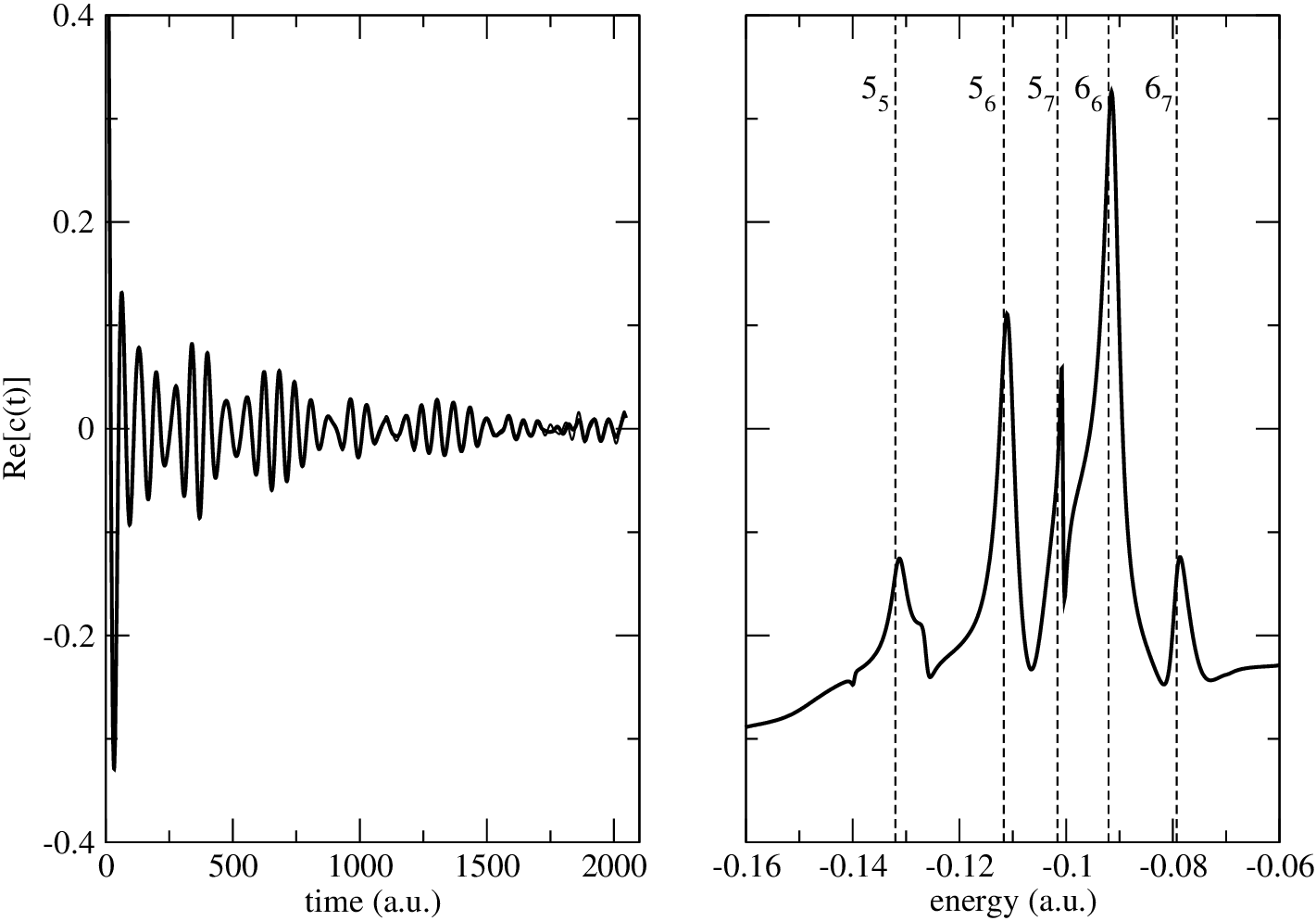}
\caption{\label{fig:resospe}
Results obtained for the $eZe$ configuration with the initial state
having parameters ${\bf q}_0^T=(30.0,30.0)$, ${\bf p}_0^T=(0,0)$ and
$\gamma_1=\gamma_2=0.1$. The left panel presents two curves for the
real part of autocorrelation function that are difficult to resolve on
the scale of the figure: a thick curve showing results obtained using
about $3.6\times 10^5$ trajectories and a thin curve showing results
obtained with about half this number of trajectories. The right panel
shows the energy spectrum obtained by the harmonic inversion
method. The vertical dashed lines indicate quantum levels.}
\end{center}
\end{figure}

Figure \ref{fig:resospe} presents further results for the $eZe$
configuration, obtained using an initial state with parameters ${\bf
q}_0^T=(30.0,30.0)$, ${\bf p}_0^T=(0,0)$, and
$\gamma_1=\gamma_2=0.1$. This state projects onto resonances which
form the lowest members of the Rydberg series with $N=5$ and
$N=6$. Well-converged calculations of the autocorrelation function
were obtained using $3.6\times 10^5$ trajectories with energies
between -4.0 and 0 a.u. No trajectories were discarded due to large
prefactors. Agreement with quantum values is seen to be very good.

Results from these and additional, similar, calculations for singlet and
triplet states of the $eZe$ configuration are displayed in Table
\ref{tab:eZe_ener}. The semiclassical values reported are the
real parts of the harmonic inversion parameters $-E_k$ of
(\ref{harminv}). Perhaps due to the reduction in the prefactor
modulus, the imaginary parts of these parameters are usually about an
order of magnitude larger than the quantum values, and are not
reported here.

\begin{figure}
\begin{center}
\includegraphics[scale=0.65]{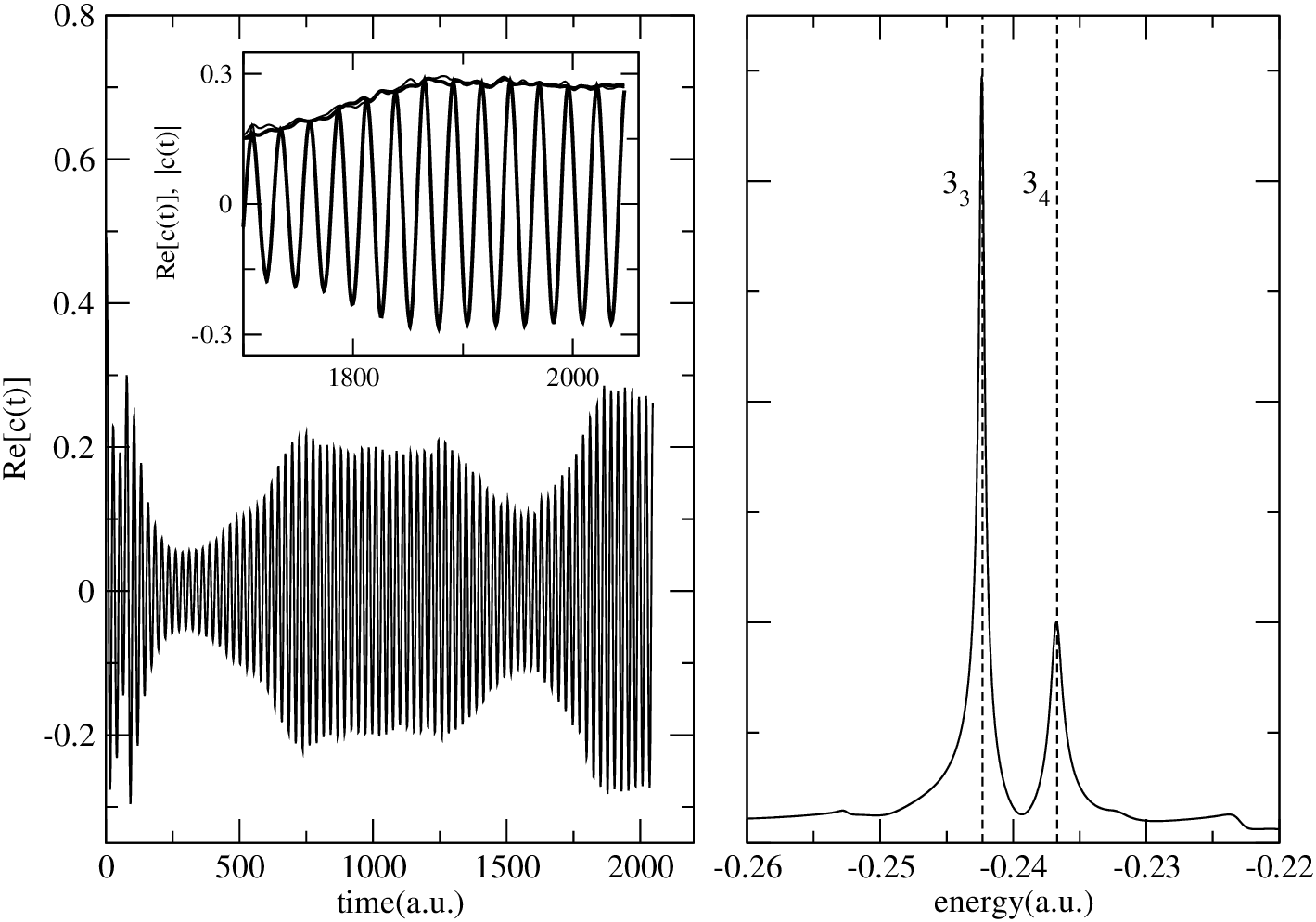}
\caption{\label{fig:zeespe}
Results obtained for the $Zee$ configuration with the initial state
having parameters ${\bf q}_0^T=(8.37,22.41)$, ${\bf p}_0^T=(0,0)$ and
$\gamma_1=\gamma_2=0.4$. The left panel shows the real part of the
autocorrelation functions. The inset shows details of the real part
and modulus of $c(t)$ obtained with about $2.3\times 10^6$
trajectories (heavy curve) and $1.2\times 10^5$ trajectories (light
curve). The right panel shows the energy spectrum obtained by the
harmonic inversion method. The vertical dashed lines indicate quantum
levels.}
\end{center}
\end{figure}

In our treatment of the $Zee$ configuration, lowest energy levels
corresponding to $n = N$ and $n=N+1$ for specific Rydberg series $N$
were calculated by choosing the $(\mathbf{q}_0,\mathbf{p}_0)$
parameters of the initial states to coincide with the location of the
central frozen planet orbit for this system. Since this orbit passes
near the phase space point ${\bf q}_0^T=(2.15,5.76)$, ${\bf
p}_0^T=(0,0)$ at $E=-1$,\cite{Richter92b} the initial state for
calculation of quantum states at an approximate energy $E$ was chosen
to have parameters ${\bf q}_0^T=(2.15/|E|,5.76/|E|)$, ${\bf
p}_0^T=(0,0)$. Figure \ref{fig:zeespe} shows results obtained in this
way with the estimate $E = -0.257$ a.u.\ and $\gamma_1=\gamma_2=0.4$. About
$2.3\times 10^6$ trajectories with energies between $-3.0$ and $0$
a.u.\ were used to produce the convergent autocorrelation curve shown
in this figure.

Results for these and other $Zee$ states, obtained in a similar way,
are reported and compared to quantum energies in Table
\ref{tab:Zee_ener}.  Since, for this configuration, our semiclassical
calculations yielded numerically identical energy levels for symmetric
and antisymmetric states with $n>N$, only one set of energies is
reported for each set of quantum numbers. It appears likely that the
singlet-triplet splitting in these cases is due to tunnelling through
the electron-electron repulsion barrier. It is known that the HK
method is unreliable for treatment of such ``deep'' tunnelling
phenomena.\cite{Kay05}

\section{Discussion and Conclusion}

Periodic orbit methods are superior tools for understanding how
quantization emerges from classical behavior. However IVR techniques
provide a way for performing semiclassical calculations that may
remain practical even when the periodic orbit treatments are
unfeasible. The results of this work show that the such methods can be
successfully applied to systems with more than one electron.

Such applications are, however, not trivial. The Coulomb singularity,
the high degree of chaos, and the instability of the classical
dynamics characterizing these systems, significantly complicate the
use of IVR methods.  Nevertheless, these difficulties can be largely
overcome, allowing calculation of converged autocorrelation signals
for times that are long enough to quantize the systems treated. Among
the methods used to achieve this were regularization of the motion,
reduction of the modulus the HK prefactor, removal of trajectories
with sufficiently negative energies, and application of the harmonic
inversion method to form the autocorrelation spectrum.

The problems mentioned in the Introduction, concerning the classical
autoionization of many-electron systems, turned out to be somewhat
less serious than anticipated. The classical escape of electrons does
require the calculation of additional trajectories, but the
computational effort involved is partly offset by savings obtained by
terminating trajectories once they ionize.

Agreement between the semiclassical results and quantum energies was
generally very good. The semiclassical energies obtained for almost
all bound and resonance states examined were within about 1\% of the
exact values. This accuracy was achieved even for the ground state of
the $eZe$ configuration. We note that previous semiclassical
treatments of this state\cite{Ezra91,Wintgen92} provide the current
semiclassical estimates of the ground state energy for the
three-dimensional atom.

In practice, even with the use of harmonic inversion, resolution of
individual levels in the energy spectrum requires calculation of the
autocorrelation function for longer times when the density of states
increases. Since, for each $N$, this density becomes infinite as $n
\to \infty$, IVR methods are not suitable for calculating levels with
large values of $n-N$. However, as the energy increases ($|E| \to 0$),
the scaling relation, (\ref{tscale}), implies that IVR
calculations can be continued for longer times without encountering
significant problems due to chaos, autoionization, etc. The results
obtained here demonstrate that this allows application of the IVR
method to determine energies for the first few resonances of each
Rydberg series $N$, even when $N$ is moderately large.

As stated in the Introduction, the present work is motivated by the
possibility of performing semiclassical calculations for the fully
three-dimensional helium atom and other atomic and molecular
systems. Since the number of trajectories needed in the present
calculations is already large, attempts to treat such systems would
certainly benefit from improvements in the efficiency of the IVR
technique. One way in which this can be accomplished is by exploiting
the energy scaling of the classical motion more thoroughly. In the
present work, separate trajectories were calculated for each physical
initial condition sampled in the four-dimensional phase space. With an
appropriate reformulation, it should be possible to achieve savings by
running trajectories at a single energy only and scaling the
results.\cite{Takatsuka06} A second way to increase the efficiency of
the calculations is to replace integration over one of the remaining
phase space variables with an integration over time along the
trajectories\cite{Elran99a,Elran99b}. Calculations for other systems
show that this approach significantly reduces the number of
independent phase space points which need to be sampled to obtain
converged results.  Still greater improvements can be often be
achieved by applying a family of IVR methods which cut the
dimensionality of the phase space integration by half.\cite{Sklarz04}

Even without these modifications, the present HK treatment may prove
to be successful for many states of the three dimensional helium
atom. Since the motion perpendicular to the collinear configurations
is known to be stable, it may perhaps be said that the present
calculations for the $eZe$ configuration already treat the most
problematic subspace of the three-dimensional system.  Thus,
application of IVR methods seem to place full-dimensional
semiclassical calculations of atoms and molecules within reach.

\section*{Acknowledgments}
The authors thank Dr. Gregor Tanner and Dr. Nenad Simonovi\'c for
useful discussions. This work was funded by the Israel Science
Foundation (Grant No. 85/03).

\appendix*
\section{Quantum calculations}

Energy levels for singlet states of the $eZe$ configuration were
previously calculated quantum mechanically by Bl\"umel and
Reinhardt\cite{Blumel92} using the complex rotation
method.\cite{Burgers95,Ho83,Reinhardt82,Moiseyev98} In addition, three
levels for this system were estimated by Simonovi\'c\cite{Simonovic04}
from a computed quantum autocorrelation spectum. To obtain values for
comparison with our calculations, which also include energies for
triplet states of the $eZe$ configuration as well as energies for
states of the $Zee$ configuration, we carried out quantum complex
rotation calculations that are described in this Appendix. The
energies obtained in this way for the singlet $eZe$ states, reported
in Table \ref{tab:eZe_ener}, agree with those of \onlinecite{Blumel92}.

Since resonance widths were not reliably obtained with the
semiclassical method investigated in this paper, no effort was made to
determine imaginary parts of resonance energy eigenvalues in the
present quantum calculations. Additionally, since singlet-triplet
splittings were not resolvable for the $Zee$ configuration by our
semiclassical method, our quantum calculations for this configuration
were limited to singlet states.

The basis set used for bound and resonance states of the $eZe$
configuration consisting of the functions
\begin{equation}
\phi_{nm}(r_1,r_2) = N_{nm}\left [(ar_1)^n(ar_2)^m  \pm
(ar_1)^m(ar_2)^n\right ] e^{-a(r_1+r_2)}, \quad n,m \geq 1,
\end{equation}
where the $+$ and $-$ signs apply for singlet and triplet states,
respectively. $N_{nm}$ is a normalization constant and
$a=|a|e^{-i\theta}$ is a complex scaling parameter. For $Zee$ singlet
states, the basis consisted of the functions
\begin{equation}
\phi_{nm}(r_1,r_2) = N_{nm}(ar_<)^n(ar_> - ar_<)^m e^{-ar_>} \quad n,m
\geq 1, 
\end{equation} 
where $r_<$ and $r_>$ denote, respectively, the smaller and larger of
$r_1$ and $r_2$. In all cases, the required overlap and Hamiltonian
matrix elements can be readily expressed in terms of elementary
functions.

To orthogonalize the basis set and remove linear dependences, the
overlap matrix was diagonalized and the Hamiltonian was represented in
the basis of the orthogonal eigenvectors associated with eigenvalues
greater than a cutoff value, typically chosen as $1.0\times 10^{-16}$.
The Hamiltonian matrix was then diagonalized with variable basis sizes
and fixed complex $a$ until convergence was obtained for the real
parts of desired eigenvalues. Typical values for the scaling
parameters were $\theta = 30$ rad and $|a|=2/N$ (for $eZe$ states) or
$|a|=1/N$ (for $Zee$ states).  Fewer than 600 basis functions were
usually required to achieve convergence for the modest precision
required here, although for some high-energy resonances, more than
1400 functions were used. $|a|$ and $\theta$ were then varied until
the real parts of the desired eigenvalues were stationary. With
sufficiently large basis sets, this condition was often already
achieved with the initial choice of these parameters so that further
optimization was unnecessary.

\newpage

\begin{table}

\caption{\label{tab:eZe_ener}
Binding energies [$-\re(E_{Nn})$] for symmetric (singlet) and
antisymmetric (triplet) states of the of the collinear helium atom in
the $eZe$ configuration. SC: values calculated with the present IVR
technique; QM: energies obtained using the quantum mechanical complex
rotation treatment described in the Appendix.}
\begin{ruledtabular}
\begin{tabular}{cccccc}
\multicolumn{2}{c}{State}&\multicolumn{2}{c}{Symmetric}&\multicolumn{2}{c}{Antisymmetric}\\
\cline{1-2} \cline{3-4} \cline{5-6}
 N & n & SC & QM & SC & QM \\
\hline
$1$& $1$ & 3.2102 & 3.2459 &  -      & -      \\
   & $2$ & 2.2225 & 2.2028 &  2.2393 & 2.2254 \\
$2$& $2$ & 0.8216 & 0.8224 &  -      & -      \\
   & $3$ & 0.6045 & 0.6098 &  0.6150 & 0.6184 \\
$3$& $3$ & 0.3621 & 0.3662 &  -      & -      \\
   & $4$ & 0.2883 & 0.2890 &  0.2906 & 0.2935 \\
   & $5$ & 0.2615 & 0.2603 &  0.2616 & 0.2618 \\
$4$& $4$ & 0.2050 & 0.2061 &   & \\
   & $5$ & 0.1690 & 0.1695 &   & \\
   & $6$ & 0.1528 & 0.1532 &   & \\
   & $7$ & 0.1437 & 0.1441 &   & \\
$5$& $5$ & 0.1314 & 0.1320 &   & \\
   & $6$ & 0.1112 & 0.1117 &   & \\
   & $7$ & 0.1006 & 0.1016 &   & \\
$6$& $6$ & 0.09127 & 0.09204 &   & \\
   & $7$ & 0.07922 & 0.07928 &   & \\
\end{tabular}
\end{ruledtabular}
\end{table}

\begin{table}
\caption{\label{tab:Zee_ener}
Binding energies for states of the collinear helium atom in the $Zee$
configuration. SC: values calculated with the present IVR technique;
QM: energies obtained using the quantum mechanical complex rotation
treatment described in the Appendix.}
\begin{ruledtabular}
\begin{tabular}{cccc}
\multicolumn{2}{c}{State}& \multicolumn{2}{c}{$-\re(E_{Nn})$}
\\
\cline{1-2} \cline{3-4}
N & n & SC & QM \\
\hline
$1$& $1$  & 2.1265 & 2.1084  \\
   & $2$  & 2.0556 & 2.0504  \\
$2$& $2$  & 0.5408 & 0.5394  \\
   & $3$  & 0.5260 & 0.5240  \\
$3$& $3$  & 0.2424 & 0.2420  \\
   & $4$  & 0.2368 & 0.2360  \\
$4$& $4$  & 0.1370 & 0.1368  \\
   & $5$  & 0.1342 & 0.1339  \\
$5$& $5$  & 0.08788 & 0.08783 \\
   & $6$  & 0.08620 & 0.08618 \\
\end{tabular}
\end{ruledtabular}
\end{table}

\newpage

\setlength{\abovecaptionskip}{25pt}
\setlength{\belowcaptionskip}{10pt} 

\end{document}